# Orientation Mapping via Dictionary Indexing of AC-STEM Kikuchi patterns using Open-source Software

Shamail Ahmed[1,2*], Johannes Haust[1,2], Felix Gruber[1,2], Celina Becker[1,2], Jürgen Belz[1,2], Andreas Beyer[1,2], Laura-Alena Schäfer[3] and Kerstin Volz[1,2*]

[1] Department of Physics, Philipps-Universität Marburg, Hans Meerwein Str. 6, 35032 Marburg, Germany

[2] Marburg Center for Quantum Materials and Sustainable Technology (mar.quest) and Department of Physics, Philipps-Universität Marburg, 35043 Marburg, Germany

[3] Institute for Ceramic Materials and Technologies (IKMT), University of Stuttgart, Allmandring 7B, Stuttgart, 70569, Germany

* Corresponding authors email: Shamail.ahmed@physik.uni-marburg.de; kerstin.volz@physik.uni-marburg.de

## Abstract:

The properties of polycrystalline materials are strongly influenced by the spatial arrangement and orientations of individual grains within the microstructure, making nanoscale characterization of grain orientation essential. This is also often the case for small grains in the nm regime explored using scanning transmission electron microscopy (STEM). Automated crystal orientation mapping (ACOM) is traditionally performed using spot-like diffraction patterns. In contrast, orientation mapping based on transmission Kikuchi diffraction (TKD) using an aberration-corrected (AC) convergent STEM probe remains relatively underexplored, despite its superior orientation sensitivity and higher spatial resolution. In this work, we present an open-source software-based template-matching approach for orientation mapping using AC-STEM TKD. A master pattern (a simulated angular distribution of Kikuchi band intensities on the unit sphere) is first generated through a dynamical simulation implemented in open-source software. This resulting pattern is subsequently imported into another open-source package for geometric simulations and orientation indexing. We demonstrate the capability of the proposed method by applying it to orientation mapping in $BaZr_{0.4}Ce_{0.4}Y_{0.1}Yb_{0.1}O_{3-\delta}$ (BZCYYb4411) fuel-cell material and $LiNiO_2$ (LNO) lithium-ion battery cathode material. The best-matched simulated patterns exhibit strong agreement with experimental data, even under the challenging conditions with limited diffraction space available for matching.

## 1. Introduction:

Scanning transmission electron microscopy (STEM) is a powerful technique for nanoscale orientation and phase analysis in polycrystalline materials.[1–5] Thereby, a spot-like nano-beam diffraction (NBD) pattern can be generated by employing a small semi-convergence angle (typically ≤ 2 mrad). By scanning this quasi-parallel probe across the sample and acquiring diffraction patterns at each scan position, a comprehensive four-dimensional (4D) dataset is produced, enabling determination of the crystal orientation and phase.[6–8] In addition, electron beam precession mitigates dynamic diffraction effects, improving the reliability of template matching with kinematic diffraction simulations.[9,10] Automated crystal orientation and phase mapping based on scanning nano-beam diffraction (SNBD) and scanning precession electron diffraction (SPED) 4D datasets have become a standard approach, with numerous commercial and open-source software packages available.[11–13] Recent advances have demonstrated the potential of machine learning for orientation mapping.[14]

In contrast to a quasi-parallel probe, a standard high-resolution convergent-beam STEM probe results in broad overlapping diffraction discs at each scan position. For relatively thick samples, these convergent-beam electron diffraction (CBED) patterns also show Kikuchi bands arising from dynamical diffraction of electrons that were first inelastically scattered and subsequently undergo coherent diffraction. Utilizing these scanning transmission Kikuchi diffraction (STKD) patterns for orientation mapping offers distinct advantages over conventional SNBD and SPED. The superior spatial resolution offered by these probes is particularly critical for characterizing small grains and narrow interfaces, such as those found in battery materials. TKD patterns also exhibit high sensitivity to crystal tilt, which significantly enhances the precision of the resulting orientation maps.[15] Unlike sparse spot patterns, the high density of intersecting Kikuchi bands provides continuous geometric constraints for the indexing algorithm. This redundancy effectively narrows the solution space, reducing the probability of mis-indexing and ensuring more robust crystallographic solutions. A significant advantage of 4D-STKD is its ease of implementation without changing the microscope's operating modes, as it utilizes the conventional convergent-beam mode standard to most modern STEM instruments. This capability proves invaluable during in situ HR-STEM experiments, where the region of interest frequently deviates from the zone-axis orientation.

While the TKD-based automated crystal orientation mapping (ACOM) indexing technique has become a standard approach for orientation mapping in the SEM, the same methodology has not achieved comparable adoption within the STEM community despite its significant inherent advantages. Several reasons may account for this limited integration. First, the TKD pattern formation is fundamentally governed by the interplay between inelastic scattering and subsequent coherent Bragg diffraction. At the lower accelerating voltages used in SEM for TKD (20–30 kV), the higher scattering cross-section increases the likelihood of the dual scattering events required for Kikuchi band formation. The resulting broader angular distribution of electrons is well-suited to TKD indexing and enables efficient orientation mapping, whereas the more forward-focused electron trajectories in high-kV (100-300 kV) STEM produce a narrower Kikuchi signal than in SEM-based transmission geometries. Secondly, the maximum achievable angular acceptance is significantly constrained in a STEM-based TKD modality compared to SEM-TKD. This limitation arises from the physical blocking of high-angle diffraction signals by the finite bore diameter of the post-specimen column. Additionally, high-angle signals near the column periphery are prone to being compromised by geometric distortions. These distortions could be introduced through the complex optical path of the intermediate and projector lenses, as well as in-column Omega-filter lenses and the TEM aberration corrector. Furthermore, the STEM-TKD field currently lacks an optimized, fully automated workflow capable of reliably indexing the restricted-angle TKD patterns characteristic of STEM.

Burton et al.[16] demonstrated the feasibility of automated STEM-TKD pattern indexing using a Hough-transform-based approach integrated with EDAX's OIM Analysis™ software. By utilizing a 150 µm condenser aperture, they achieved an enhanced signal-to-noise ratio; however, this came at the expense of spatial resolution, which was limited to approximately 25 nm. Additionally, the reliance on commercial software suites may present accessibility challenges for researchers seeking open-source or highly customizable workflows. Additionally, the approach using the Hough transformation to detect Kikuchi bands in STEM diffraction patterns exhibits some disadvantages. In this approach, taking into account the detector-sample geometry calibration, band positions are converted into plane normals and compared with crystallographic reference angles to index the bands and determine crystal orientation.[17] Although Hough-based indexing is widely used, its performance degrades when band contrast is low or the signal-to-noise ratio is poor. Indexing is also challenging for low-symmetry and multiphase materials, where pseudo-symmetry and structurally similar phases can lead to ambiguities.[18] It often struggles to distinguish similar crystal structures with slightly different lattice parameters, as it relies primarily on bands' central positions and occasionally on integrated band intensity.[19] In STEM, the reduced acceptance angle further limits its effectiveness, especially when only a few strong reflections are captured, and weak reflections dominate the pattern.

Template matching, i.e., matching experimental TKD/EBSD patterns with the simulated ones, which are simulated including dynamical electron scattering effects (commonly referred to as dictionary indexing (DI)), is a powerful and established technique in the SEM community in EBSD and TKD modalities. It helps mitigate some of the challenges of the Hough approach.[20–22] In contrast to feature extraction, DI bypasses the need for explicit feature extraction from the TKD patterns. Instead, experimental patterns are matched against a comprehensive library of templates generated via physics-based forward modeling. This approach demonstrates superior robustness in high-noise experimental patterns,[23] a critical advantage for STEM-TKD. In the case of STEM-TKD, reliable band detection via the Hough transform is often hindered by noise at high-angle scattering and the absence of major zone axes due to restricted diffraction space (which is often further reduced by the necessity of masking the central bright-field disk).

In this study, an open-source computational workflow is employed to simulate TKD patterns, accounting for dynamical electron scattering effects at 200 kV. These simulations serve as templates for matching against experimental 4D STKD datasets acquired via an aberration-corrected STEM probe at high spatial resolution. Master patterns are generated using the EMsoft software package[18,24,25] and subsequently imported into Kikuchipy,[26,27] which is an open-source Python-based library. The geometrical simulations in accordance with the detector geometry are performed utilizing TKD master patterns, and these geometrically simulated patterns are matched with experimental patterns. The findings confirm the feasibility of indexing STKD patterns with a highly restricted angular diffraction space. Using the proposed open-source pipeline, we achieved robust orientation solutions with high confidence levels, successfully overcoming the limitations typically encountered in TEM-based TKD.

We applied template matching using STKD datasets to two diverse materials: a proton-conducting perovskite oxide (BZCYYb4411) and a lithium-ion battery cathode active material (LNO). The orientation mapping results obtained from these investigations demonstrate that the methodology presented here serves as a powerful complementary technique to conventional SNBD and SPED orientation mappings.

## 2. Methods:

This section describes the procedures for acquiring the dataset and processing the patterns. The open-source software used, along with its key parameters and settings, are then introduced.

### 2.1. Dataset acquisition and pattern processing:

The 4D STEM-TKD datasets were acquired in aberration-corrected STEM convergent-beam mode using a double-aberration-corrected JEOL 2200/FS S/TEM operated at 200 kV. Data acquisition was performed with a TVIPS TemCam-XF416 CMOS camera controlled by the PentaRhei software (CEOS GmbH), using dwell times of 53 ms (for the BZCYYb4411 sample) and 60 ms (for the LNO sample) per pixel through the REVELON TEM Scan Controller from point electronic GmbH. The diffraction patterns were acquired using the full camera sensor area in 8x binning mode, resulting in images of 512x512 pixels, which were subsequently binned to 128x128 pixels. The camera length was adjusted to provide a total angular extent of 15.8° on the full screen of the camera along horizontal and vertical dimensions. This diffraction pattern calibration was used for all the datasets in the paper. The datasets were initially stored in HyperSpy's native hspy format and subsequently converted to Kikuchipy's native .h5 format after pattern processing. Kikuchipy is built on HyperSpy's data structures and signal framework, extending them with EBSD-specific functionality.[26,28,29] This integration enables efficient and streamlined data handling from acquisition through processing and indexing. Before indexing, the dataset must be corrected for any beam shift (descan) to ensure accurate pattern center (PC) values for geometric simulations and template matching.

Figure 1 presents the pattern preprocessing workflow for a Si [110] FIB sample. Figure 1(a) shows the raw TKD pattern with the zone axis slightly tilted from the exact [110] orientation. The Si [110] sample is presented here to illustrate the pattern processing steps, since its diffraction pattern is familiar to those working in TEM. Figure 1(b) presents the same pattern after logarithmic scaling. Figure 1(c) shows the band-pass filtered image. Band-pass filtering was applied to the log-scaled STEM-TKD patterns in the frequency domain using a Gaussian filter (standard deviation σ = 5 pixels, implemented via the built-in Kikuchipy function) to suppress slowly varying background intensity (low-frequency components) while preserving the spatial frequencies associated with Kikuchi bands. The filtered patterns were obtained by dividing the original patterns by the estimated background. Subsequently, adaptive histogram equalization (AHE) was applied (using the built-in Kikuchipy function) to enhance band contrast, as shown in Figure 1(d). The pattern was then corrected for known parabolic distortions introduced by the post-sample column's lenses. Finally, the bright-field disc was masked before template matching, as illustrated in Figure 1(f).

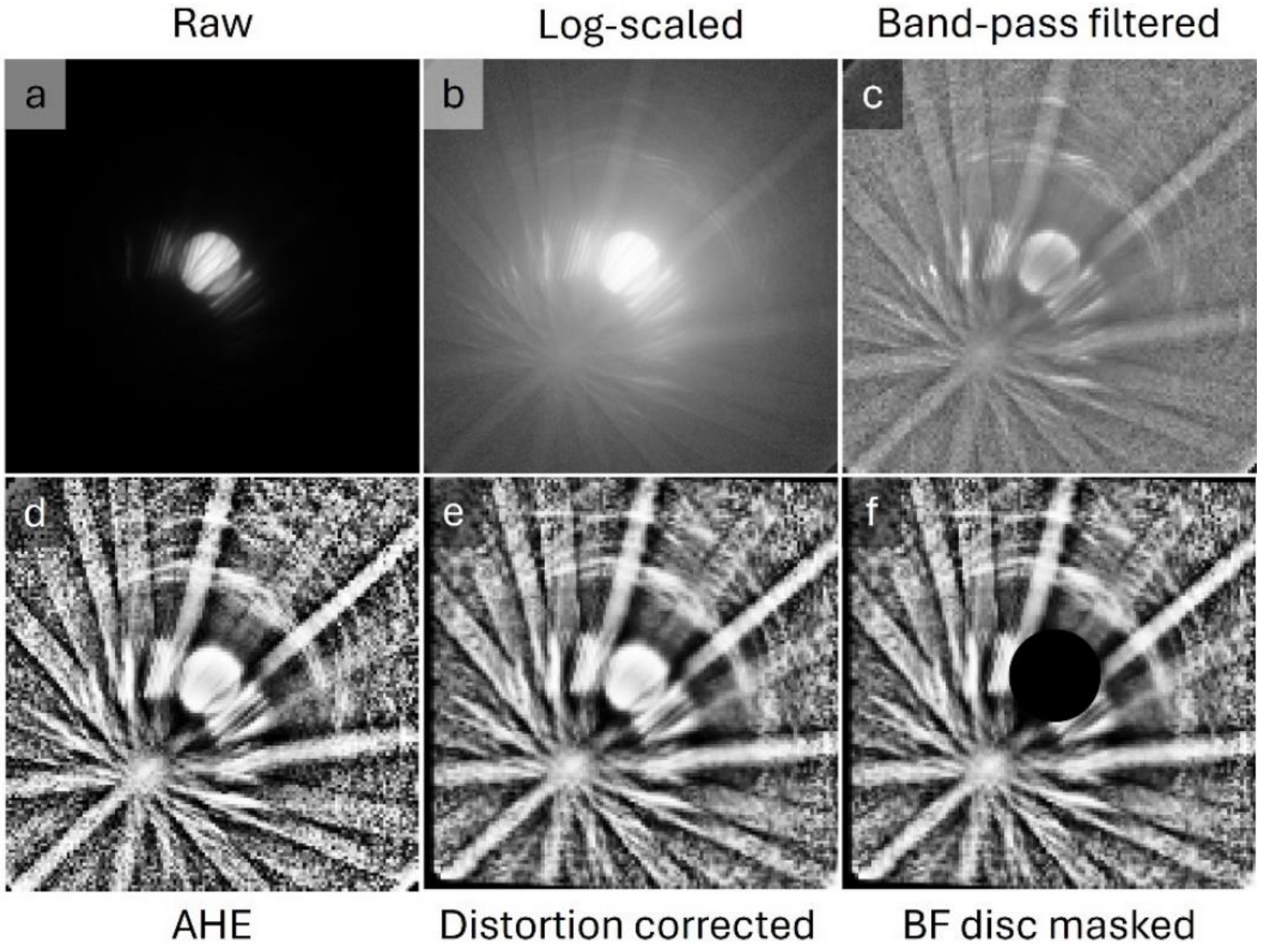

**Figure 1: Pattern Pre-processing.** Pre-processing workflow for a Si[110] TKD pattern. (a) Raw acquisition, (b) log-scaled, (c) band-pass filtered, (d) adaptive histogram equalization (AHE), (e) distortion-corrected, and (f) center-masked to exclude the bright-field disc.

### 2.2. Indexing:

After the pre-processing, the 4D STEM-TKD datasets are stored in Kikuchipy's native .h5 format for indexing. For the dynamical simulation of the TKD master pattern, an open-source EMsoft software is used. The full tutorial for the master pattern simulations is explained elsewhere.[18,30] To facilitate dictionary indexing, we first performed Monte Carlo (MC) simulations. This is carried out to compute the energy, depth, and directional distributions of scattered electrons. Subsequently, we generated TKD master patterns. A master pattern is the simulated angular distribution of Kikuchi band intensities on the unit sphere for a given crystal structure, representing the complete diffraction signal independent of detector geometry or crystal orientation. These master patterns, calculated via dynamical electron scattering theory, serve as the source data from which the orientation-specific templates are derived, based on the detector geometry, to match the experimental patterns. It should be noted that standard MC and TKD master pattern simulation parameters were employed, as detailed in the Supplementary Information. While an exhaustive exploration of the simulation parameter space remains beyond the scope of this proof-of-concept study, it is anticipated that further optimization, specifically by aligning simulation parameters with experimental conditions such as specimen thickness, will significantly enhance matching results. This potential for refinement is addressed later in the paper.

The Kikuchi pattern indexing workflow was implemented using the Kikuchipy Python library, starting with the import of experimental 4D-STEM TKD datasets and the corresponding EMsoft TKD master patterns. It should be noted that dictionary indexing can also be performed directly within EMsoft. However, Kikuchipy was selected due to its seamless compatibility with the acquired data in the .hspy format. As a Python-based package, Kikuchipy offers greater ease of implementation, as well as enhanced flexibility through its customizable functions. Its Python foundation also facilitates broader accessibility and adoption within the scientific community and enables straightforward integration with complementary Python libraries, such as orix, for efficient handling, analysis, and post-processing of results. It is also important to note that there exists a Python interface called pyEMsoft, which provides access to selected Fortran data structures and subroutines.[31] After data loading using Kikuchipy, the master pattern was projected into the Lambert azimuthal equal-area projection at an accelerating voltage of 200 kV. For dictionary indexing, a set of orientations was generated by sampling the fundamental zone using a cubochoric grid with an angular resolution of 0.6° (for the fuel cell sample) and 0.8° (for the $LiNiO_2$ sample).[32] The resolution of the sampling in the orientation space for the dictionary ultimately decides the precision of the misorientation in the resulting orientation maps. However, depending on the crystal structure and symmetry, the number of templates could increase significantly when lowering the angular resolution value. Afterwards, a virtual TKD detector was defined using calibrated pattern center (PC) coordinates (PCx, PCy, detector distance) of the camera to project the master pattern onto the camera plane and construct a simulated diffraction dictionary according to the detector (camera) geometry. In Kikuchipy, the term detector distance (DD) is used to denote the distance between the sample and the detector. The DD is defined as the ratio between the physical distance from the sample to the detector plane and the detector height, making it a dimensionless quantity equivalent to the camera length.

Finally, the experimental patterns were matched to this simulated library using a normalized cross-correlation (NCC)[33] metric within the dictionary indexing framework, resulting in a high-resolution orientation map. The dictionary generation and indexing stages were optimized through a parallelized framework to handle the high-dimensional data efficiently. By employing a chunk-based processing strategy (implemented in the Kikuchipy), simulated patterns were partitioned into manageable blocks to minimize memory overhead. This allowed for an iterative matching routine where experimental patterns were compared against the orientation library in parallel batches. A simplified workflow for the procedure adopted is presented in the schematic shown in Figure 2.

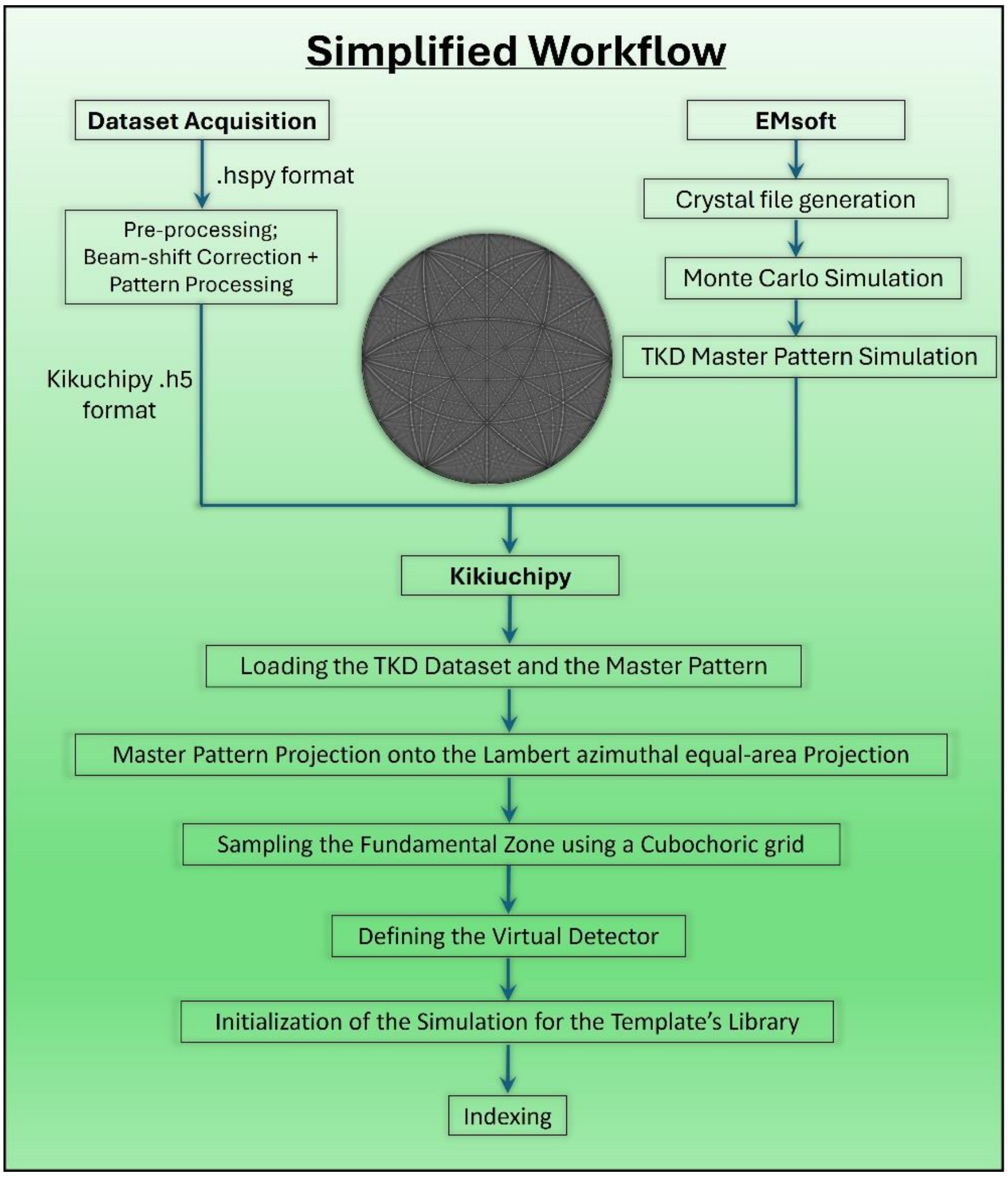

**Figure 2: Workflow.** A schematic showing the simplified workflow used for STKD dataset indexing.

It is important to mention that while dictionary indexing of STEM TKD patterns follows the same general framework as SEM-based on-axis TKD, several simulation parameters may require careful consideration when operating at high accelerating voltages. Due to the significantly smaller angular capture area inherent to STEM TKD, the indexing may rely more heavily on higher-order, weaker reflections observed at high kV. This necessitates the use of a smaller minimum d-spacing cutoff (dmin) during master pattern simulation to ensure adequate representation of these reflections. Furthermore, the reduced angular field of view implies that the NCC similarity metric becomes more sensitive to small orientation deviations, suggesting that both the master pattern resolution and the angular sampling density of the orientation dictionary may need to be increased relative to parameters conventionally used in the case of SEM on-axis TKDs. The simulation parameters adopted in the present work, as detailed in the supplementary information, are found to perform well for the datasets reported here and may serve as a practical starting point for further optimization. However, the systematic optimization of these parameters, including the appropriate dmin, master pattern pixel resolution, and cubochoric sampling density for STEM TKDs, has not been performed in the present work and remains the subject of future study.

### 2.3. Averaged-NCC Score (ANS) Maps:

Often, it is desired to visualize the internal structure of the grains before even performing indexing, as indexing via template matching can be computationally expensive. This requires a map that is sensitive to the subtle changes in the diffraction at each pixel with the surrounding. For this purpose, we define an averaged-NCC score (ANS) map. First, the NCC scores were calculated separately between the experimental diffraction pattern at each pixel (x,y) and those of its immediate horizontal (x+1,y) and vertical (x,y+1) neighbors. These two NCC values were then averaged and assigned as the intensity value of the corresponding pixel (x,y). Before the NCC calculation, a circular mask was applied to all diffraction patterns in the dataset to suppress the influence of the transmitted beam. The mask's radius was chosen to be approximately 1.5 times larger than that shown in Figure 1(f), effectively minimizing forward-scattering intensity contributions. This procedure produces a spatial map of local structural similarity. By excluding the transmitted beam, the resulting map enhances subtle variations in the diffraction signal, thereby improving the visibility of grain boundaries and crystalline defects. This metric can be mathematically defined as,

$$ANS(x,y) = \frac{1}{2}[NCC(P_{x,y}, P_{x+1,y}) + NCC(P_{x,y}, P_{x,y+1})]$$

Where ANS(x, y) is the averaged NCC score at pixel (x, y), and $P_{x,y}$ represents the bright-field disk-masked experimental diffraction pattern acquired at coordinate (x, y). Unlike the standard Averaged Dot Product (ADP) map,[18] which is also sensitive to absolute intensity variations, this local correlation map employs NCC-score comparison to focus strictly on structural similarity. By masking the direct beam and utilizing the Pearson correlation coefficient, the method provides a contrast mechanism that is invariant to linear brightness shifts, resulting in a more robust delineation of grain boundaries based on diffraction rather than the signal's amplitude. The resulting map is conceptually analogous to the correlation coefficient mapping developed by Kiss et al. for the spot-like patterns containing 4D datasets.[34]

## 3. Results:

We apply the template matching approach for orientation mapping using STEM-TKD datasets to samples BZCYYb4411 and LNO. The subsequent discussion reports our findings, followed by an analysis of the method's advantages, limitations, and potential applications. The details regarding the dictionary generation for the two samples are provided in the experimental section.

### 3.1. $BaZr_{0.4}Ce_{0.4}Y_{0.1}Yb_{0.1}O_{3-\delta}$ (BZCYYb4411) proton-conducting perovskite-type oxide:

BZCYYb4411 is a perovskite-type proton-conducting oxide that has been investigated as an electrolyte for protonic ceramic fuel and electrolysis cells.[35,36] Figure 3(a) shows a virtual high-angle annular dark-field (HAADF) image reconstructed from a 4D STEM-TKD dataset, highlighting a grain boundary between two grains of BZCYYb4411. The dataset consists of 100x100 pixels acquired with a step size of 3 nm per pixel. Figure 3(b) presents the inverse pole figure (IPF) along the x direction overlaid with the corresponding NCC score map, while the individual IPFx and NCC maps are shown in Figures S1(a) and S1(b), respectively. Figures 3(c) and 3(e) display the experimental TKD patterns extracted from points 1 and 2 in Figure 3(b). In these panels, the upper half shows the single TKD pattern at the indicated point, whereas the lower half shows the average of 16 surrounding patterns to show a better comparison with the simulated patterns. Figures 3(d) and 3(f) show the best-matching simulated patterns corresponding to the experimental patterns in Figures 3(c) and 3(e), respectively. Red arrows in Figures 3(c & e) highlight excess-deficiency features that are not captured by the simulations (Figures 3(d & f)); this aspect is discussed further later in the paper. Comparison of Figures 3(c) and 3(e) with Figures 3(d) and 3(f) demonstrates that, despite the relatively noisy patterns and limited diffraction space, template matching produces high-quality indexing results with correct orientations.

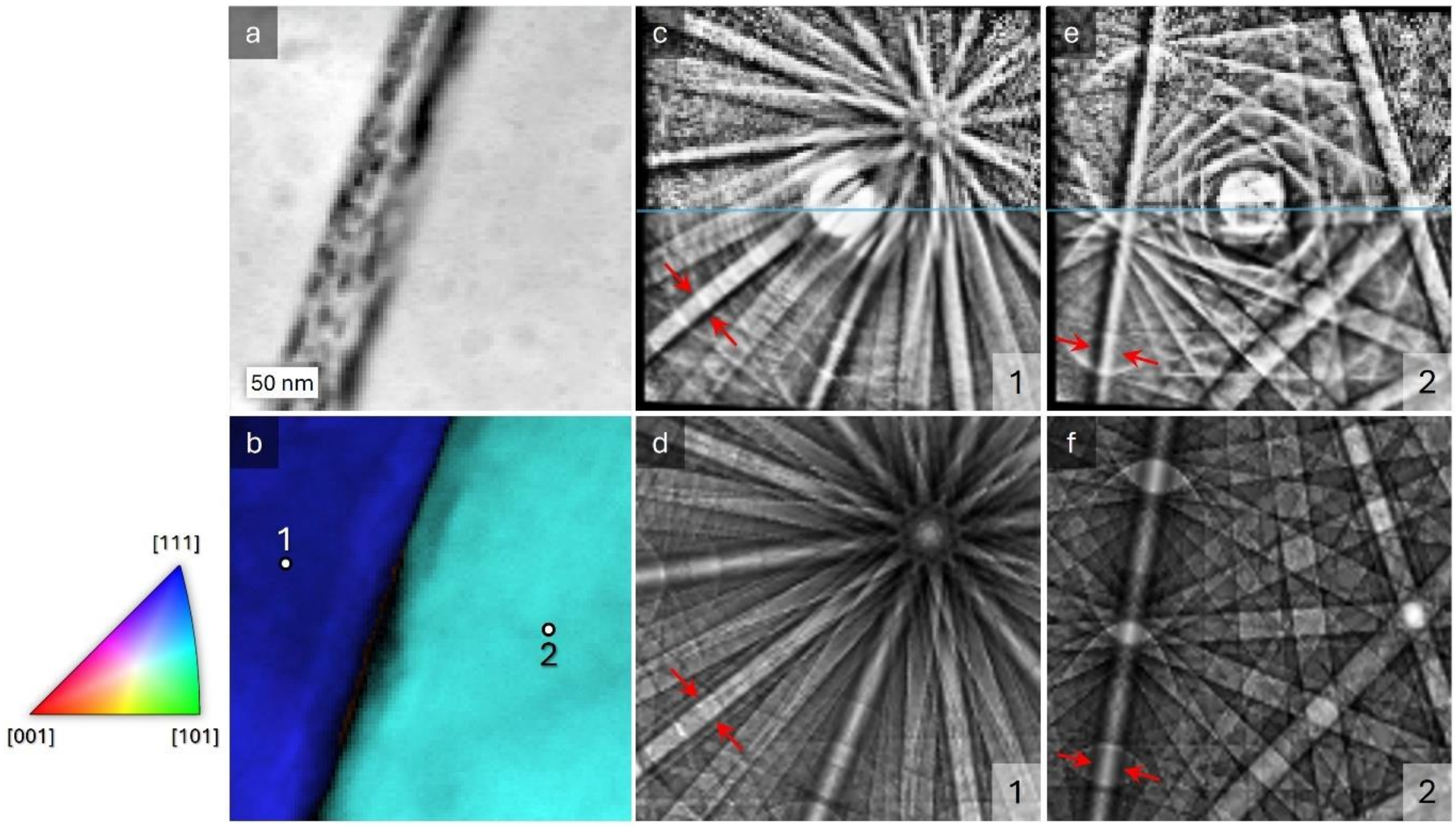

**Figure 3: Template-matching results for the BZCYYb4411 sample.** (a) Virtual HAADF image of the analyzed region. (b) Overlay of the IPFx and the NCC map. (c, e) Experimental TKD patterns extracted from points 1 and 2 in (b); the upper halves show single patterns, while the lower halves show the average of 16 surrounding patterns. (d, f) Corresponding best-matching simulated patterns for the experimental patterns in (c) and (e), respectively. The red arrows in (c) & (e) show excess-deficiency lines, while these arrows in similar positions in (d) & (c) show the absence of such features.

### 3.2. $LiNiO_2$ (LNO) lithium-ion battery (LIB) cathode active material (CAM):

LNO is one of the most important and extensively studied cathode active materials (CAMs) for lithium-ion batteries (LIBs).[37] It offers one of the highest practically achievable specific capacities in the LIBs layered CAM class, but suffers from rapid capacity fade during extended electrochemical cycling.[38] The material used in this study was annealed at 700 °C after synthesis. This treatment, after synthesis, is known to produce defects.[39] Figure 4(a) shows a virtual HAADF image reconstructed from a 4D STEM-TKD dataset, highlighting a grain boundary between two LNO primary particles. The dataset comprises 100x100 pixels with a step size of 2.5 nm per pixel. Figure 4(b) presents the IPFx overlaid with the corresponding NCC score map, while the individual IPFx and NCC maps are shown in Figures S2(a) and S2(b), respectively. Figures 4(c) and 4(e) display the experimental TKD patterns extracted from points 1 and 2 in Figure 4(b). In these figures, the upper-left halves show the single TKD patterns at the indicated points, whereas the lower-right halves show the average of 16 surrounding patterns. Figures 4(d) and 4(f) show the corresponding best-matching simulated patterns for the experimental patterns in Figures 4(c) and 4(e), respectively.

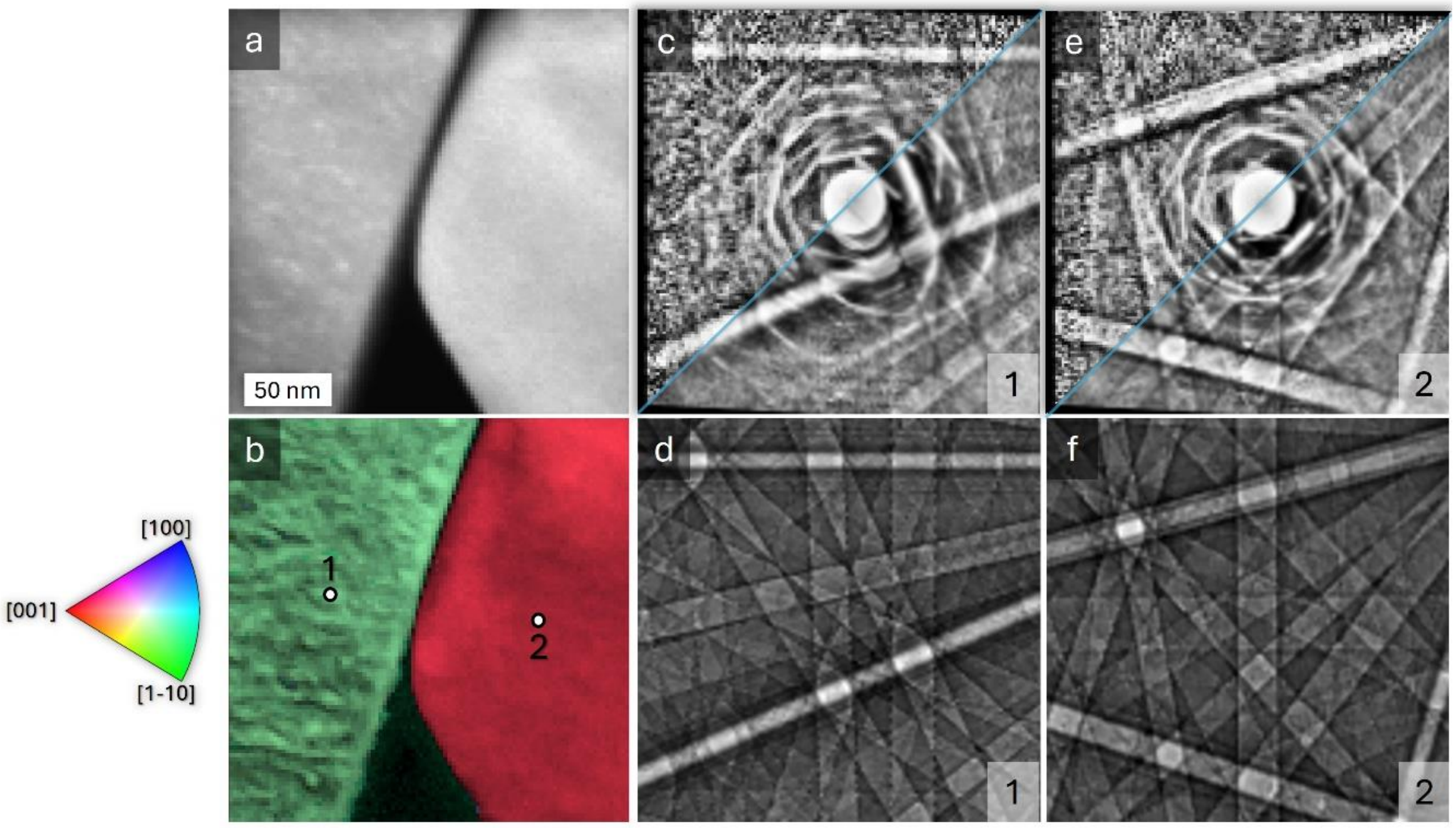

**Figure 4: Template-matching results for the LNO sample.** (a) Virtual HAADF image of the analyzed region. (b) Overlay of the IPFx and the NCC map. (c, e) Experimental TKD patterns extracted from points 1 and 2 in (b); the upper-left halves show single patterns, while the lower-right halves show the average of 16 surrounding patterns. (d, f) Corresponding best-matching simulated patterns for the experimental patterns in (c) and (e), respectively.

Comparison of Figures 4(c) and 4(e) with Figures 4(d) and 4(f) demonstrates that, despite the noisy patterns and limited diffraction space, template matching yields good matches. It's important to note that the noise level is considerably higher than the individual TKD patterns shown in the upper panels of Figures 3(c) and 3(e) for the BZCYYb4411 sample, and most of the bands in the LNO dataset are quite weak.

### 3.2.2 Characterization of subtle crystal distortions and internal structure of the grains:

It is interesting to note that the grain on the left in Figure 4(a), colored green (close to the [1-10] zone axis), exhibits irregular dark stripes in the NCC maps (also visible in Figure S2(a)), indicating regions with a sudden drop in NCC scores. Such features are absent in the neighboring grain. To investigate the origin of the dark stripes further, Figure 5(a) shows an overlay of the IPFx and NCC map, with three points marked along one of these dark stripes. Figures 5(b-d) present the TKD patterns from these points, with the 011 and 014 bands marked for reference. Across the dark stripe, the 014 plane shifts slightly toward the 011 band, as seen in Figures 5(b) and 5(c), and comes back to its position in Figure 5(d). This causes a sudden drop in the NCC score since the template from the distorted crystal is not present in the templates' library. Since the 014 band is sensitive to defects and strain, and our previous study indicates that annealing at 700 °C induces strain and defects, this behavior is consistent with expectations. This shows the sensitivity of the technique presented in this study. Because of the high-resolution aberration-corrected STEM probe together with the sensitivity of the template matching, the NCC map can produce a quite accurate representation of the internal features that could be overlooked with conventional imaging due to mixed contrast of various electron scattering phenomena, as in the HAADF image shown in Figure 4(a). This highlights the sensitivity of the technique presented in this study.

Before indexing, the internal structure of the grains can already be assessed using the ANS map as described in Section 2.3. Figure 5(e) shows the ANS map revealing the internal variations within individual grains. The dark stripes observed in Figure 5(b), which are attributed to strain and crystallographic defects, are also clearly visible in the ANS map (Figure 5(e)), highlighting the usability of such mapping.

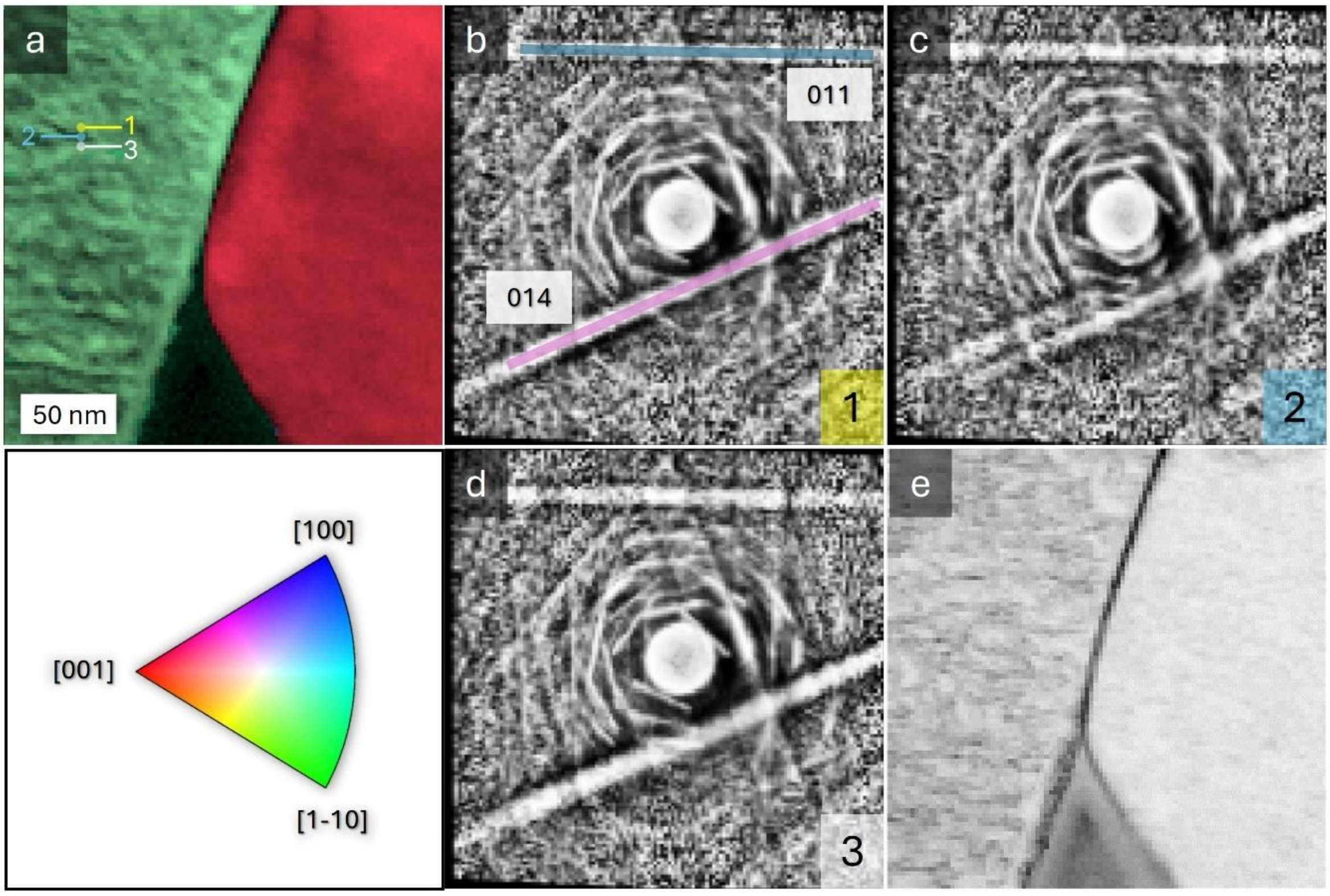

**Figure 5: TKD patterns across the dark bands in the LNO sample.** (a) Overlay of IPFx and NCC map in the LNO sample. (b-d) are the TKD patterns from points 1-3 shown in (a). The 011 and 014 bands are marked as examples in (b). (e) shows an averaged NCC score (ANS) map.

## 4. Discussion:

Although the indexing results are already highly promising, further improvements are expected as the simulated templates more closely approximate the experimental patterns. This can be achieved by incorporating realistic specimen thickness in the simulations, using models that account for excess-deficiency effects, dynamical diffraction of an incoherent diffuse intensity distribution, and properly including intensity variations associated with forward scattering in and around the bright-field disc.[40] Winkelmann et al.[41] developed a model that accounts for dynamical electron diffraction from coherent point emitters, dynamical diffraction of incoherent diffuse intensity, excess-deficiency effects arising from background intensity gradients within the Bragg angle range of Kikuchi bands, and anomalous absorption. They demonstrated that incorporating these effects significantly improves EBSD indexing performance. Later, Zhang et al.[40] showed potential strategies to implement these effects for on-axis SEM-TKD pattern simulations. The interference of coherent diffraction discs at high semi-convergence angles with Kikuchi bands, together with high intensity around the bright field (BF) disc, leads to the formation of complex features and creates significantly inhomogeneous intensity variations at the band edges, especially after background correction and subsequent adaptive histogram equalization, which are not well represented in our simulations. These features occupy a significant fraction of our angular field-of-view restricted STEM TKD patterns, as can be seen around BF discs in Figures S3 and S4. It would be valuable to investigate whether the improvements from Zhang et al.[40] (to include such features for the case of SEM on-axis TKD) can also be adapted for STEM-TKD patterns in future studies to simulate these complex features for improved dictionary indexing.

Since the Kikuchi band signals can be weak at high kVs, especially at low sample thicknesses, the implementation of multi-exposure diffraction pattern fusion, analogous to the methodology described by Zhang et al.[42] for SEM on-axis TKD can be explored as a viable pathway for signal enhancement in TEM at high scattering angles. By integrating multiple frames, the resulting patterns may exhibit significantly improved clarity and better-defined band edges, facilitating more accurate template matching. However, this may come at the cost of a higher electron-beam dose per pixel. An alternative strategy for signal enhancement involves oversampling the real space using a high-speed direct electron detector and subsequent averaging of the neighboring patterns. By acquiring twice or four times the number of TKD patterns in both x and y dimensions and subsequently performing binning in the real space, the resulting averaged patterns would exhibit a significantly improved SNR while maintaining the necessary information.

Given that dictionary-based indexing is inherently computationally intensive, various strategies to optimize processing speeds should be considered. Varley et al.[43] demonstrated that a PCA-based dimensionality reduction approach for EBSD can yield significantly higher indexing rates while maintaining comparable orientation quality in noisy datasets. Furthermore, spherical indexing techniques warrant investigation as a more computationally efficient alternative to traditional dictionary matching.[44] Finally, the development of robust deconvolution strategies is essential for resolving and indexing superposed signals originating from overlapping crystals.

STEM-TKD technique demonstrates significant potential as a robust phase-mapping tool, particularly for differentiating phases with high structural similarity, owing to its exceptional sensitivity to subtle crystallographic variations. Compared to the spot-based diffraction patterns obtained via SNBD and SPED, the intricate networks and crossings of Kikuchi lines offer much richer structural textures, significantly improving the sensitivity required to distinguish between similar phases in the STEM TKD-based technique. However, as illustrated in Figure 5, the sensitivity of the NCC score to minor deviations from the reference template is quite drastic. This implies that the phase assignment should not be based exclusively on the highest absolute NCC scores of the best-matching orientations from the candidate phases. A more reliable phase-mapping protocol would require a comparative analysis that should go beyond simple absolute NCC score-based comparisons. The construction of a metric analogous to the phase confidence metric developed by Ram et al.[19] could be explored in this regard. Furthermore, once the best-matching orientations from the candidate phases are obtained, a more efficient high-level texture-specific cross-correlation metric could be explored to decide between the best solutions. In this regard, advanced deep learning architectures, such as Convolutional Neural Networks (CNNs), Vision Transformers (ViTs), and Swin Transformers (Swin-T), hold significant potential. By leveraging their inherent ability to extract hierarchical and spatial feature representations, these models can move beyond pixel-to-pixel intensity comparisons, offering a more robust and noise-tolerant framework for phase discrimination. This will be the subject of future studies.

## 5. Conclusion:

In contrast to the well-established SNBD and SPED methodologies, orientation mapping via STEM-TKD remains an underutilized technique despite its superior sensitivity to subtle crystallographic misorientations. Unlike spot-based patterns, TKD patterns show rich textured features that have the potential to be better suited to distinguish between phases with subtle crystallographic differences, which are a common occurrence in battery materials and are often difficult to discern with spot-based patterns solely.[45,46]

This study demonstrates a robust open-source pipeline, integrating EMsoft and Kikuchipy, capable of generating dynamical simulations and performing high-fidelity template matching on the dataset acquired with an aberration-corrected convergent STEM probe with high spatial resolution. Our results for BZCYYb4411 and LNO systems confirm that experimental patterns exhibit strong correlation with their simulated counterparts. Our analysis of the LNO sample showed that, by combining the high-resolution probe of AC-STEM with the sensitivity of template matching, subtle crystallographic variations can be highlighted. Furthermore, ANS mapping can be used as a robust, computationally inexpensive method to track the subtle crystallographic distortions. While presented here as a proof-of-concept, we anticipate that further refinements in noise suppression, simulation accuracy, and computational efficiency may establish STKD as a standard tool for high-resolution orientation analysis.

## 6. Experimental Section

### Dictionary generation for the BZCYYb4411 sample

Initial SNBD characterization indicated that our BZCYYb4411 adopts a cubic structure with space group $Pm\bar{3}m$. Using a cubochoric grid with an angular resolution of 0.6°, a total of 4,325,557 templates were simulated using the master pattern and then subsequently matched with each experimental pattern in the dataset.

### Dictionary generation for the LNO sample

LNO has a layered rhombohedral structure (space group ($R\bar{3}m$). Using a cubochoric grid with an angular resolution of 0.8°, a total of 6,904,775 templates were simulated using the master pattern. This is significantly larger than the BZCYYb4411 sample at 0.6° resolution because of the cubic structure. The simulated patterns were subsequently matched with each experimental pattern in the dataset.

## Acknowledgements:

We gratefully acknowledge support from the German Research Foundation (DFG) in the framework of the research unit „Synergetisches Design protonenleitender Keramiken für Energietechnologie (SynDiPET) (project number 556363981). Funding from the European Regional Development Fund (ERDF) and the Recovery Assistance for Cohesion and the Territories of Europe (REACT-EU) is gratefully acknowledged.

## Declaration for the Use of AI

Large language model chatbots, such as ChatGPT and Gemini, were used solely for paraphrasing, language enhancement, grammatical correction, and improving the coherence of the original text. These tools were not used to generate new ideas or original academic content. All concepts, analyses, and conclusions presented in this work are the authors' own.

# Supplementary Information:

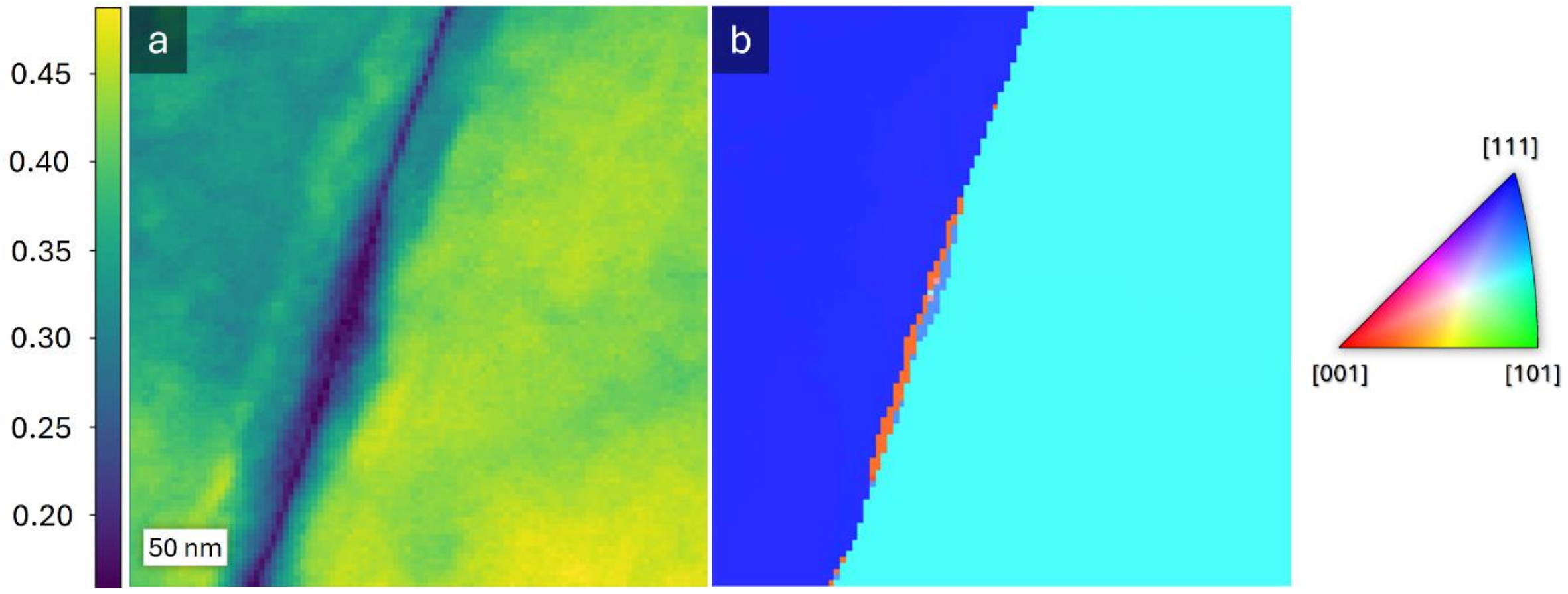

**Figure S1:** (a) NCC score map and (b) is the corresponding IPFx of the region shown in Figure 3(a).

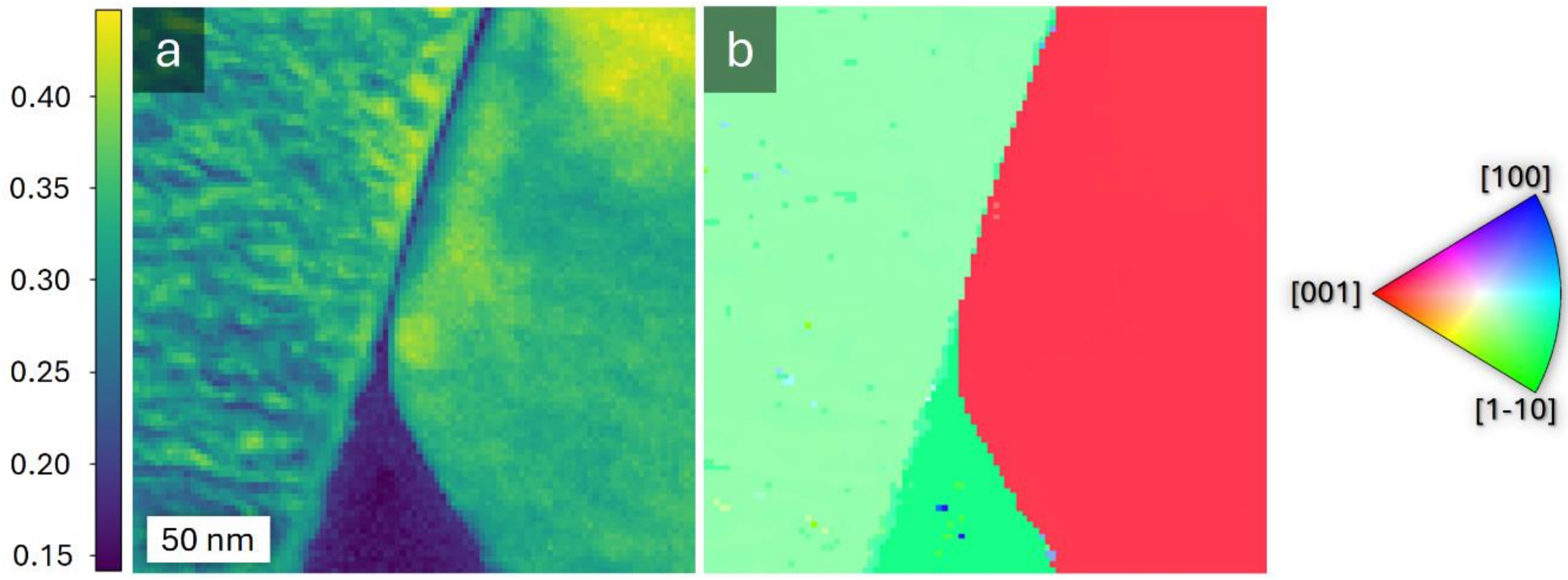

**Figure S2:** (a) NCC score map and (b) is the corresponding IPFx of the region shown in Figure 4(a).

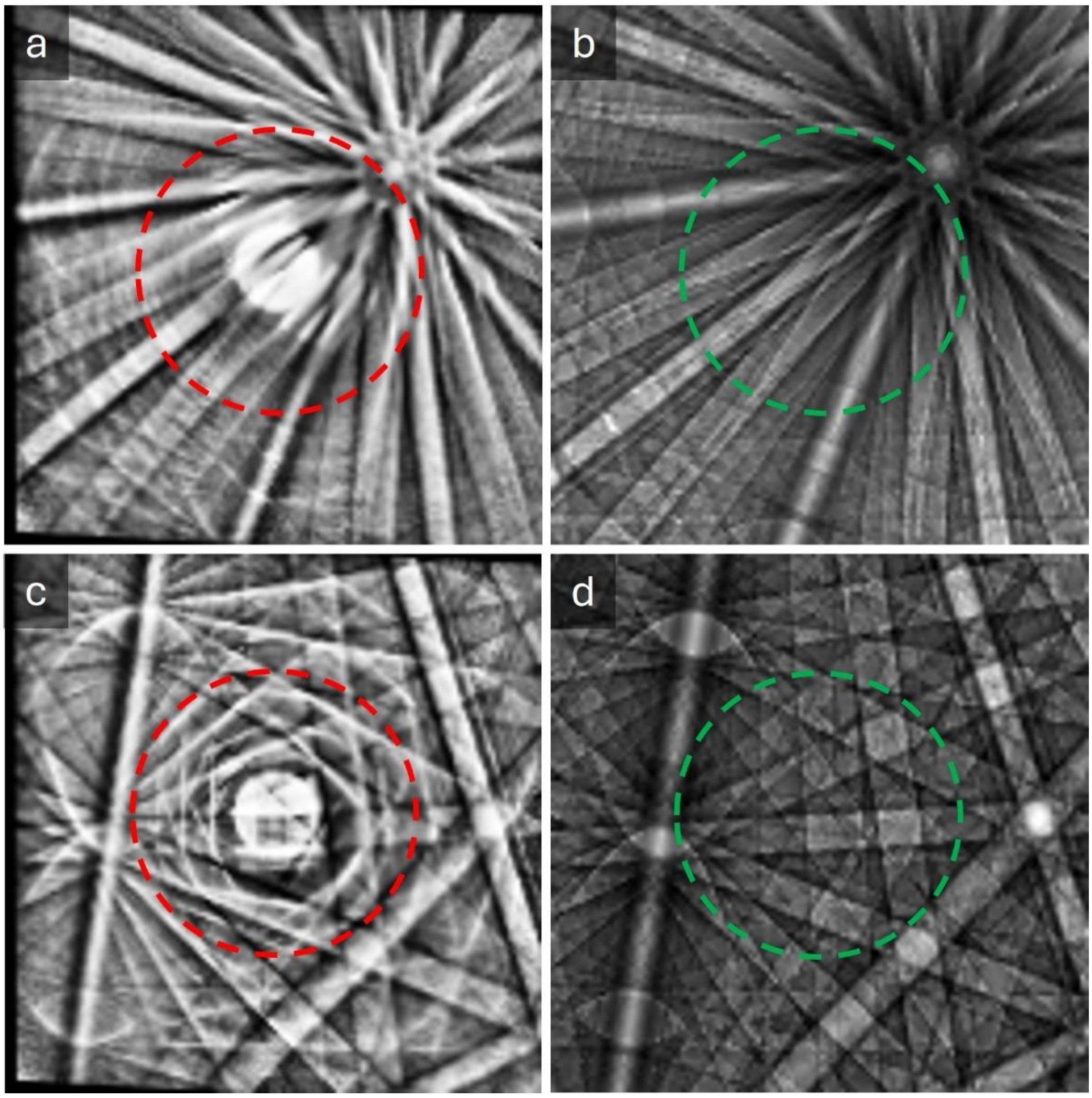

**Figure S3:** (a & c) are averaged diffraction patterns from two different grains of the BZCYYb4411 sample, as shown in Figure 3, and (b & d) are their corresponding best matching simulated patterns. The dashed red circles in the a & c and their corresponding green dashed circles in the b & d, respectively, show the regions that do not match between the experiments and the simulations.

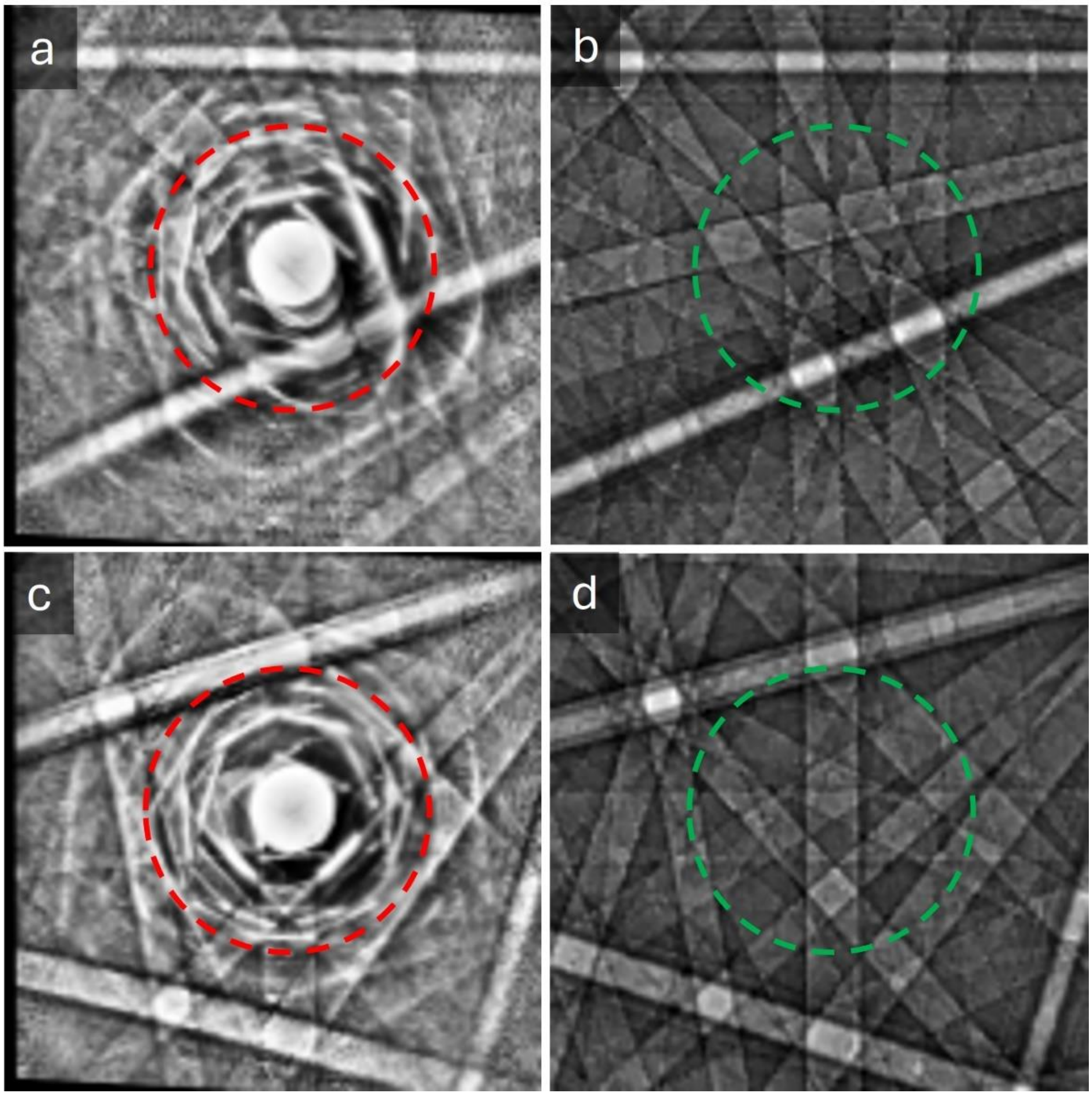

**Figure S4:** (a & c) are averaged diffraction patterns from two different grains of the LNO sample, as shown in Figure 4, and (b & d) are their corresponding best matching simulated patterns. The dashed red circles in the a & c and their corresponding green dashed circles in the b & d, respectively, show the regions that do not match between the experiments and the simulations.

The details about different Monte Carlo and TKD master pattern simulation parameters can be found at https://github.com/EMsoft-org/EMsoft/wiki/.

### Monte Carlo simulation parameters for BZCYYb4411 (EMsoft)

```
&MCCLfoildata
! name of the crystal structure file
xtalname = 'BZCYYb4411.xtal',
! for full mode: sample tilt angle from horizontal [degrees]
sig = 0.0,
! sample tilt angle around RD axis [degrees]
omega = 0.0,
! number of pixels along x-direction of square projection [odd number!]
numsx = 1001,
! number of incident electrons per thread
num_el = 10,
! GPU platform ID selector
platid = 1,
! GPU device ID selector
devid = 1,
! number of work items (depends on GPU card; leave unchanged)
globalworkgrpsz = 150,
! total number of incident electrons and multiplier (to get more than 2^(31)-1 electrons)
totnum_el = 2000000000,
multiplier = 1,
! incident beam energy [keV]
EkeV = 200.D0,
! minimum energy to consider [keV]
Ehistmin = 197.D0,
! energy binsize [keV]
Ebinsize = 1.0D0,
! max depth [nm] (this is the maximum distance from the bottom foil surface to be considered)
depthmax = 100.0D0,
! depth step size [nm]
depthstep = 1.0D0,
! total foil thickness (must be larger than depth)
thickness = 200.0,
! output data file name; pathname is relative to the EMdatapathname path !!!
dataname = 'BZCYYb4411_MCoutput.h5'
/
```

### TKD Master Pattern simulation parameters for BZCYYb4411 (EMsoft)

```
&TKDmastervars
! smallest d-spacing to take into account [nm]
dmin = 0.05,
! number of pixels along x-direction of the square master pattern  (2*npx+1 = total number)
npx = 1000,
! name of the energy statistics file produced by EMMCfoil program; relative to EMdatapathname
! this file will also contain the output data of the master program
energyfile = 'BZCYYb4411_MCoutput.h5',
! number of OpenMP threads
```

```
 nthreads = 15,
! setting this parameter to .TRUE. forces the program to add all atom contributions together
! instead of keeping them separate, which is the default; this option can be used to keep the
! master file for structures with many atoms in the asymmetric unit to a manageable size.
 combinesites = .FALSE.
! restart computation ?
 restart = .FALSE.,
! create output file with uniform master patterns set to 1.0 (used to study background only)
 uniform = .FALSE.,
 /
```

## Monte Carlo simulation parameters for LNO (EMsoft)

```
&MCCLfoildata
! name of the crystal structure file
xtalname = 'LNO.xtal',
! for full mode: sample tilt angle from horizontal [degrees]
sig = 0.0,
! sample tilt angle around RD axis [degrees]
omega = 0.0,
! number of pixels along x-direction of square projection [odd number!]
numsx = 1001,
! number of incident electrons per thread
num_el = 10,
! GPU platform ID selector
platid = 1,
! GPU device ID selector
devid = 1,
! number of work items (depends on GPU card; leave unchanged)
globalworkgrpsz = 150,
! total number of incident electrons and multiplier (to get more than 2^(31)-1 electrons)
totnum_el = 2000000000,
multiplier = 1,
! incident beam energy [keV]
EkeV = 200.D0,
! minimum energy to consider [keV]
Ehistmin = 197.D0,
! energy binsize [keV]
Ebinsize = 1.0D0,
! max depth [nm] (this is the maximum distance from the bottom foil surface to be considered)
depthmax = 120.0D0,
! depth step size [nm]
depthstep = 1.0D0,
! total foil thickness (must be larger than depth)
thickness = 150.0,
! output data file name; pathname is relative to the EMdatapathname path !!!
dataname = 'LNO_MCoutput.h5'
/
```

## TKD Master Pattern simulation parameters for LNO (EMsoft)

```
&TKDmastervars
! smallest d-spacing to take into account [nm]
dmin = 0.015,
! number of pixels along x-direction of the square master pattern  (2*npx+1 = total number)
npx = 1000,
! name of the energy statistics file produced by EMMCfoil program; relative to EMdatapathname
! this file will also contain the output data of the master program
energyfile = 'LNO_MCoutput.h5',
! number of OpenMP threads
nthreads = 14,
! setting this parameter to .TRUE. forces the program to add all atom contributions together
! instead of keeping them separate, which is the default; this option can be used to keep the
```

```
! master file for structures with many atoms in the asymmetric unit to a manageable size.
 combinesites = .FALSE.
! restart computation ?
 restart = .FALSE.,
! create output file with uniform master patterns set to 1.0 (used to study background only)
 uniform = .FALSE.,
 /
```